\documentclass[floatfix,twocolumn,showpacs,prl,
preprintnumbers]{revtex4}
\usepackage{amssymb}
\usepackage{epsfig}
\usepackage{graphicx}
\usepackage{dcolumn}
\usepackage{bm}

\begin{document}

\title{Complete spin polarization of electrons in semiconductor layers and
quantum dots}
\author{V.~V.~Osipov$^{1,2}$, A.~G.~Petukhov$^{3}$, 
and  V.~N.~Smelyanskiy$^{2}$}
\date{\today }
\affiliation{$^{1}$New Physics Devices, LLC, 2041 Rosecrans Avenue, El Segundo, CA 90245 
\\
$^{2}$NASA Ames Research Center, Mail Stop 269-3, Moffett Field, CA 94035 \\
$^{3}$Physics Department, South Dakota School of Mines and Technology, Rapid
City, SD 57701 }
\date{\today}

\begin{abstract}
We demonstrate that non-equilibrium electrons in thin nonmagnetic 
semiconductor layers or quantum dots can be fully spin polarized by
means of simultaneous electrical spin injection and extraction. The
complete spin polarization is achieved if the thin layers or quantum
dots are placed between two ferromagnetic metal contacts with moderate
spin injection coefficients and antiparallel magnetizations.  
The sign of the spin polarization is determined by the
direction of the current. Aplications of this effect in spintronics
and quantum information processing are discussed.
\end{abstract}

\maketitle
Spintronics is a new field of condensed matter physics based on manipulation
of electron spins is solids \cite{Zut,Aw}.
%An idea for the manipulation of electron spin in solids has led to the
%development of a new field: spintronics \cite{Zut,Aw}. 
Injection of
spin-polarized electrons into nonmagnetic semiconductors (NS) is of
particular interest because of the relatively large spin-coherence lifetime, 
$\tau _{s}$, in NS \cite{Zut,Aw} and promising applications for ultrafast,
low-power spin devices \cite{Zut,Aw,Datta,Hot,OBO} and for spin-based quantum
information processing (QIP) \cite{Aw,QIP}.
Both the characteristics of the
spintronic devices and fidelities of various spin-based QIP schemes
dramatically 
improve when spin polarization $P_{n}$ 
of electrons in NS is high, i.e. when $P_{n}\rightarrow
100\% $. 
Thus the main challenge is to
achieve  high values of $P_n\sim 100\%$ in NS  both at
room and  low temperatures.   A simple intuition, 
supported by all previous theoretical works 
\cite{Aron,Mark,Rash,Flat,BO,OB}, suggests that $P_{n}$ under electrical spin
injection cannot exceed either
the
polarization of a spin source or the spin injection coefficient of a
ferromagnet-semiconductor contact. Since conventional ferromagnetic 
metals (FMs) rootinely used
in semiconductor technology have moderate spin polarizations $P_{n}\lesssim $
40\%, the main strategy to improve spin injection has been identified
as development of magnetic
semiconductors and half-metals with high bulk spin polarizations \cite{Zut,Aw}.
Nonetheless, the greatest value of electron spin polarization in NS 
to date, $P_{n}\simeq $
32\%, has been observed in spin injection experiments with conventional FM (Fe)
contacts \cite{Jonk}.

One of the obstacles for the spin injection from FM into NS is a high and
wide potential Schottky barrier that normally forms in NS near
FM-NS interfaces \cite{sze}. The spin injection from FM into NS
corresponds to a reverse current in the Schottky junction, which is usually
negligible. Therefore, a thin heavily doped $n^{+}$-S layer between FM and NS
has been used to increase the tunneling transparency of the barrier 
\cite{sze} and the spin injection current \cite{OB,BO,Jonk}. Since the spin
injection is  a tunneling of spin polarized electrons from FM into NS,
which is a symmetric process, the spin selection should also occur in
forward-biased FM-S junctions when electrons are emitted from NS into FM \cite%
{BO}. In this case the spins are extracted from NS into FM and the 
tunneling FM-$n^{+}$-S contact works as a spin filter.

In this letter we consider a donor doped semiconductor ($n$-NS) layer placed
between two spin selective contacts, for example, tunneling FM-$n^{+}$-NS
junctions, Fig.1. We show that the electron spin polarization in this $n$-NS
layer %\ and quantum dots 
can achieve 100\% even when the spin injection
coefficients of the contacts, $\gamma _{L}$ and $\gamma _{R}$, are quite
moderate. The only requirement is that $\gamma_{L(R)}$ should
 weakly depend on the current, $J$. Indeed, the currents of up-electrons
with spin $\sigma =\uparrow $ through the contacts can be written as 
\begin{equation}
J_{\uparrow }(0)=J(1+\gamma _{L})/2,\text{ }J_{\uparrow }(w)=J(1+\gamma
_{R})/2,  \label{bc}
\end{equation}%
where $J_{\uparrow }(0)$ and $J_{\uparrow }(w)$ are the currents in the $n$%
-NS layer at the boundaries with the right and left contacts  $x=0$ and $%
x=w$ (Fig.1).  The continuity equation reads \cite%
{Aron,Flat,OB,BO} 
\begin{equation}
dJ_{\uparrow }/dx=q\delta n_{\uparrow }/\tau _{s},  \label{eq}
\end{equation}%
where\ $\delta n_{\uparrow }=n_{\uparrow }-n_{S}/2$, $n_{S}$ is the total
electron density and $n_{\uparrow }$ is the density of up-electrons. Obviously 
$\delta n_{\uparrow }(x)\simeq const$ in the $n$-NS layer when its thickness $%
w\ll L_{s}$, where $L_{s}$ is the spin diffusion length.
Integrating Eq.~(\ref{eq}) over $x$ from $0$ to $w$, we obtain $J_{\uparrow
}(w)-J_{\uparrow }(0)=qw\delta n_{\uparrow }/\tau _{s}$, and  using (%
\ref{bc}) and $\delta n_{\downarrow }(x)=-\delta n_{\downarrow }(x)$ we find
the spin polarization in thin $n$-NS layer:   
\begin{equation}
P_{n}(x)=(\delta n_{\uparrow }-\delta n_{\downarrow })/n_{S}\simeq (\gamma
_{R}-\gamma _{L})J/J_{w} \equiv {\bar P}_n,  \label{pn}
\end{equation}%
where \ $J_{w}=qn_{S}w/\tau _{s}$.  One can see
from (\ref{pn}) that: $P_{n}=0
$ when $\gamma _{R}=\gamma _{L}$, since in this case the currents of up- and
down electrons through the right and left contacts are the same: $%
J_{\uparrow }(0)=J_{\uparrow }(w)$. On the contrary when $\gamma _{R}\neq
\gamma _{L}$ the value $\left\vert P_{n}\right\vert \rightarrow 1$ when
 $J\rightarrow J_{t}$ $=J_{w}/(\gamma _{R}-\gamma _{L})$. For
example, if $\gamma _{L}<0$ and $\gamma _{R}>0$, i.e. magnetizations M$_{1}$
and M$_{2}$ in FM have opposite directions, Fig.1(a), $P_{n}=(\left\vert
\gamma _{R}\right\vert +\left\vert \gamma _{L}\right\vert )J/J_{w}$, i.e. $%
P_{n}\rightarrow 1$ as $J\rightarrow J_{t}=J_{w}/(\left\vert
\gamma _{R}\right\vert +\left\vert \gamma _{L}\right\vert )$. Thus, due to
the difference of the currents  of spin polarized electrons through the
contacts, the spin density $P_{n}$ in the $n$-NS layer increases with the
current and can reach $100\%$ even at small spin injection coefficients. One
can see from (\ref{pn}) that the inversion of the current  results
in the opposite sign of $P_{n}$. These findings are valid  for both
degenerate and nondegenerate semiconductors.
\begin{figure}[htbp]
\includegraphics[width=.77\linewidth,clip=true]{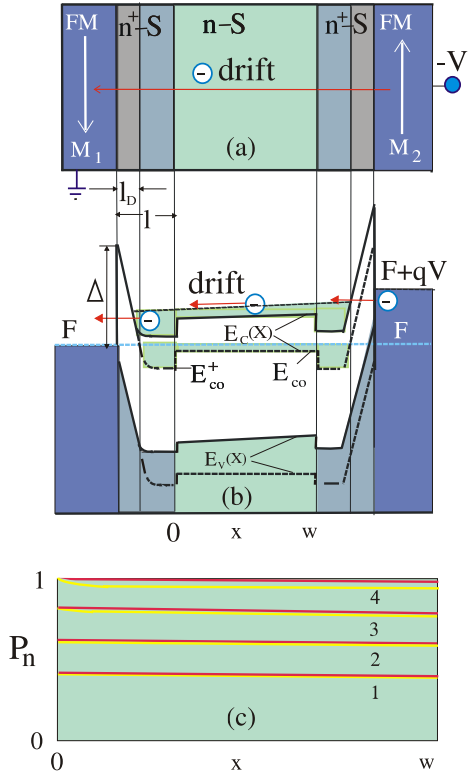}
\caption{(a) Ferromagnet-semiconductor FM-$n^{+}$-$n$-$n^{+}$-FM
heterostructure containing a donor doped nonmagnetic semiconductor 
($n$-NS)
layer sandwiched between two ferromagnetic metals (FM) having opposite
magnetizations, M$_{1}$ and M$_{2}$, and  two thin layers of a heavily
doped degenerate semiconductor (n$^{+}$-S layers) between the $n$-NS 
and FM
layers. (b)
 Energy diagrams of the heterostructure in equilibrium (broken
curves) and at  bias voltage $V$ (solid curves) 
in the case when the $n^{+}$%
-$S$ layers have a narrower bandgap than that of the $n$-$S$ region
and the $n$-$S$ region is a degenerate semiconductor 
(for the nondegenerate $%
n$-$S$ region $E_{c0}>F$). Here $F$ is the Fermi level in equilibrium, $%
E_{c}(x)$\ is bottom of semiconductor conduction band, $E_{c0}$ and $%
E_{c0}^{+}$ are the values of $E_{c}(x)$ in the $n$-S and $n^{+}$-S layers,
respectively; $w$ and $l$ thicknesses of the $n$-S and n$^{+}$-S layers,
respectively; $\Delta $ and $l_{D}$ are the height and thickness of the
Schottky barrier of the FM-$n^{+}$-S junctions. (c) Spatial dependence
 of electron spin
polarization, $P_{n}(x)$, in the nondegenerate (red curves) and
degenerate semiconductor $n$-NS 
layer (yellow curves) for $\protect\gamma %
_{R}=-\protect\gamma _{L}=0.3$ and $w=0.1L_{s}$ at currents $J=0.7J_{w}$
(curves 1), $J=J_{w}$ (curves 2), $J=1.4J_{w}$ (curves 3), $%
J=J_{t}=0.99J_{w}/(P_{J}^{R}-P_{J}^{L})=1.65J_{w}$ (curves 4). Red curves
are calculated from Eq.(\protect\ref{PN}) and yellow curves are obtained
from
numerical analysis of the diffusion-drift equation for $\protect\delta %
n_{\downarrow }(x)$.}
\end{figure}

In nondegenerate NS the diffusion constants and mobilities of up- and
down-electrons are the same, $D_{\uparrow }=D_{\downarrow }=D$ and $\mu
_{\uparrow }=$ $\mu _{\downarrow }=$ $\mu $, and the distribution of $\delta
n_{\downarrow }(x)$ for any $w$ reads \cite{Flat,OB,BO}: 
\begin{equation}
\delta n_{\uparrow }(x)=(n_{S}/2)(c_{1}e^{-x/L_{1}}+c_{2}e^{-(w-x)/L_{2}}),
\label{n}
\end{equation}%
where $L_{1,2}=L_{s}\left( \sqrt{1+(J/2J_{S})^{2}}-J/2J_{S}\right) $, $L_{s}=%
\sqrt{D\tau _{s}}$, $J_{S}=qn_{S}L_{s}/\tau _{s}$, and $q$ is the elementary
charge. Using the boundary conditions (\ref{bc}) we obtain: 
\begin{eqnarray}
P_{n}(x) = \frac{J}{J_{S}}\left[ \frac{L_{s}(\gamma _{R}-\gamma
_{L}e^{w/L_{2}})}{L_{1}\left( e^{w/L_{2}}-e^{-w/L_{1}}\right) }e^{-x/L_{1}}
\right .%
 +  \nonumber \\
\left . \frac{L_{s}(\gamma _{R}-\gamma _{L}e^{-w/L_{1}})}{L_{2}\left(
e^{w/L_{2}}-e^{-w/L_{1}}\right) }e^{x/L_{2}} \right]  \label{PN}
\end{eqnarray}%
It follows from (\ref{PN}) that $P_{n}(0)$ or $P_{n}(w)$ reaches 1 at
$J=J_{t}(\gamma_L,\gamma_R,w)$. The spatial dependence of 
$P_{n}(x)$  is very weak for $w\ll L_{s}$ and $P_n(x)\simeq {\bar P}_n$
as in Eq.~(\ref{pn}).

One can see from (\ref{pn}) and (\ref{PN}) that $\gamma _{R}$ and $\gamma
_{L}$ determine a particular value of the threshold current, $J_{t}$, 
but it does not alter the
main result: $\left\vert P_{n}\right\vert \rightarrow 1$ as $J\rightarrow
J_{t}$. The only requirement is a relatively weak dependence of $\gamma _{R}$
and $\gamma _{L}$ on $J$. This condition can be fulfilled, for example, when
thin, heavily-doped $n^{+}$-S layers are formed between  $n$-NS and FM
regions, Fig.~1. The parameters of  $n^{+}$-S layers have to satisfy 
certain conditions \cite{OSP}. In particular, the electron gas must be
degenerate in a certain part of the $n^{+}$-S layer and the transition
between the $n^{+}$-S and $n$-NS layers must be step-like. This situation
is realized when the $n^{+}$-S layers have energy bandgaps  narrower than that
of the $n$-S region, Fig.~1(b), or when an additional $\delta$-doped 
acceptor layer
is formed between the $n^{+}$-S and $n$-NS regions. Due to a high
density of electrons in the $n^{+}$-S layer, $\gamma$ of such FM-$n^{+}$-S
contacts weakly depends on $J$ up to the currents significantly
exceeding $J_{t}$ \cite%
{OSP}. Weak dependence of $\gamma $ on $J$ is also realized  in 
FM-NS junctions 
with highly degenerate  accumulation layers formed in NS near
the FM-NS interface. This situation has been studied extensively 
in Refs. \cite%
{Aron,Mark,Rash,Flat} and has been realized expeimentally in Fe-InAs junctions
\cite{Ohno}.

We note that $\gamma $ \cite{Remk} can strongly depend on $J$ \cite%
{BO,OB} in FM-NS junctions when the $n$-NS region and thin 
 ($\delta -$doped) $n^{+}$%
-layer  are both nondegenerate semiconductors. In these
FM-NS-FM junctions 
%with $\delta -$doped layers 
$P_{n}$ attains the value  $%
P_{n}=(\gamma _{R}-\gamma _{L})/(1-\gamma _{R}\gamma _{L})$ for  $J>J_{S}$
and $w\ll L_{1,2}$ \cite{OBO}. One can see that $P_{n}=0.6$ at $\gamma
_{R}=-\gamma _{L}=0.33$, i.e. even in this case $P_{n}$ can exceed the spin
injection coefficients of the FM-S junctions.

In degenerate semiconductors the diffusion constants of up- and
down-electrons are different and depend on electron densities. As a result, $%
\delta n_{\downarrow }(x)$ is described by a diffusion-drift equation with a
bi-spin diffusion constant $D(P_{n})$ which goes to zero when $\left\vert
P_{n}\right\vert \rightarrow 1$ \cite{OSP}. Therefore,
in degenerate NS, the spatial 
variation of $P_{n}(x)$ is 
sharper and its current dependence is stronger 
 than those in nondegenerate NS [cf. yellow and
red curves in Fig.~1(c)]. For example, at $w=0.1L_{s}$ and currents $J=J_{t}$
the polarization changes in the ranges $1>P_{n}(x)>0.984$ or $%
1>P_{n}(x)>0.954$ \ for nondegenerate NS and $1>P_{n}(x)>0.91$ or $%
1>P_{n}(x)>0.85$ for degenerate NS when $\gamma _{R}=-\gamma _{L}=0.3$ or $%
\gamma _{R}=-\gamma _{L}=0.1$, respectively.

At high currents when $J>J_{t}$ [see e.g. (\ref{pn}) and \ref{PN}]
the value $\left\vert P_{n}\right\vert =2\left\vert \delta n_{\downarrow
}\right\vert /n_{S}=\left\vert 2n_{\downarrow }-n_{S}\right\vert /n_{S}$
is higher than 1, i.e. spin density either near $x=0$ or near   $x=w$
exceeds $n_{S}$. This  means that the used condition of neutrality $%
n_{\uparrow }+$ $n_{\downarrow }=n_{S}$, i.e. $\delta n_{\downarrow
}(x)=-\delta n_{\downarrow }(x)$, is violated and a negative space charge
accumulates near one of the boundaries.
% of  $n$-NS region with  $n^{+}$%
%-S layer. 
This charge will increase the voltage drop  $%
V_{S}$ across $n$-NS region. This conclusion  
follows from numerical analysis of our system of equations which includes:
Eq.~(\ref%
{eq}), $J=J_{\uparrow }(x)+J_{\downarrow }(x)=\mathrm{const}$,
$J_\sigma=q\mu n_\sigma E + q D_\sigma
{dn_\sigma}/{dx}$,
and the Poisson's
equation, $\varepsilon \varepsilon _{0}dE/dx=q(n_{s}-n_{\uparrow }-$ $%
n_{\downarrow })$.

The FM-$n^{+}$-$n$-$n^{+}$-FM heterostructures under consideration are
supersensitive spin valves. Indeed, as we noticed above [see (\ref{pn}) and (%
\ref{PN})], $P_{n}\simeq 0$ and $\delta n_{\downarrow }(x)\simeq 0$ when $%
w<L_{s}$ and $\gamma _{R}=\gamma _{L}$, i.e. when the magnetizations M$_{1}$
and M$_{2}$ in the FM layers have the same direction. In this case, $V_{S}$ $%
=V_{ohm}=Jw/q\mu n_{S}$ at any current. When $\gamma _{R}\neq \gamma _{L}$
the space charge arises in the $n$-S region at $J>J_{t}$ and $V_{NS}$
exceeds $V_{ohm}$ significantly. 
Thus, inversion of direction of M$_{1}$or M$%
_{2}$ and also precession of electron spin in the $n$-S region have to
change the voltage $V_{S}$ to a much greater extent than in 
the structures considered
in Refs. \cite{OBO}. The use of the FM-$n^{+}$-$n$-$n^{+}$-FM
heterostructures in different spin-based devices \cite{Zut,Aw,Datta,Hot,OBO}
should result in dramatic improvement
 of their characteristics
due to the high spin polarization of electrons in the $n$-NS layer.

\begin{figure}[htb]
\includegraphics[width=.9\linewidth,clip=true]{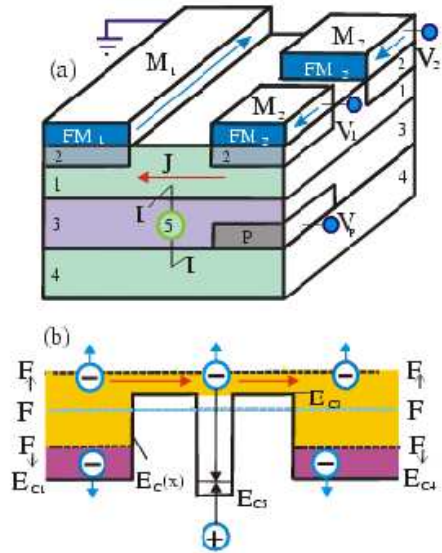}
\caption{Schematic drawing of a ferromagnet-semiconductor heterostructure
(a). FM$_{1}$ and FM$_{2}$ are ferromagnetic layers with magnetization 
M$_{1}$ and M$_{2}$ (blue arrows); chlorine (1,4,5) 
and cerise (3) areas are
donor doped semiconductor regions with 
different band-gap energy $E_{g}$ and
placement of bottom of conduction 
band $E_{c}$: $E_{c3}>E_{c1}=E_{c4}\gtrsim
E_{c5}$, respectively; light blue areas (2) are heavily doped
degenerate semiconductor (n$^{+}$-S) 
regions with $E_{c2}<E_{c1}$; light
grey area P is acceptor doped $p$-semiconductor region. 
(b). Energy diagrams
of the heterostructure in sections I-I shown in Fig.2(a). $F$ is the
equilibrium Fermi level, $F_{\uparrow }$ and $F_{\downarrow }$ are quasi
Fermi levels for up- and down-electrons, respectively.}
\end{figure}

A more complex FM-$n^{+}$-$n$-$n^{+}$-FM heterostructure is shown in
Fig.~2(a). High spin polarization of electrons arises in a thin $n$-S layer 1
grown on highly resistive widebandgap semiconductor layer 3 when the current
flows between ferromagnetic  regions FM$_{1}$ and FM$_{2}$. Such a
heterostructure can be utilized  modernize various spin-based nanodevices
proposed in Refs. \cite{OBO}
such as ultraspeed magnetic sensors,\ transistors, square law detectors, and
frequency multipliers. It can also be used for
100\% spin polarization of electrons in quantum dots 5 localized in the
layer 3 grown on a semiconductor substrate 4 having the same bandgap as
the layer 1. This effect can be realized at very 
low temperatures even when
the spin polarization inside the $n$-S layer 1 is less then 100\%, e.g. $%
P_{n}\simeq $60\%-80\%. Indeed, let us consider quantum dots (QD) (regions
(5) in Fig.~2(a) with band offset $\Delta E_{c}^{/}=E_{c5}-E_{c3}<0$)
that are placed inside of 
a two-dimensional potential barrier (semiconductor layer
(3) with band offset $\Delta E_{c}=E_{c3}-E_{c1}>0$). 
As we noted above, the
spin polarization $P_{n}$ and density of up-electrons, $n_{\uparrow }$,
increase with current $J$ between FM$_{1}$and FM$_{2}$ contacts when $\gamma
_{L}<0$ and $\gamma _{R}>0$. This means that the quasi-Fermi level for the
up-electrons $F_{\uparrow }$ increases and $F_{\downarrow }$ for
down-electrons decreases with $J$. At a certain 
current $F_{\uparrow }$ can
exceed $\Delta E_{c}$ and only the up-electrons will populate the 
layer 3 and
will be captured by QDs. Thus, at very low temperatures the electron
polarization in the QDs should be extremely close to 1, 
$1-P_{n}\simeq \exp
[-(E_{c3}-F_{\downarrow })/T]\ll 1$. 
The spin polarization of electrons\ in
QDs can be realized after their recombination with photogenerated holes or
holes injected from p-region, shown in Fig.2a. This effect can be used for
efficient polarization of nuclear spins in QDs. The sign of the spin
polarization\ of electrons in the $n$-NS layer (1) and in QDs can be reversed
by simple inversion of the current direction between FM$_{1}$and FM$_{2}$
contacts, Fig.~2(a). Thus, the considered effects have a potential for
changing the development strategy for various spin-based devices and QIP
schemes.

This work is supported by NASA ITSR program (V.~O and V.~S.) 
and NSF (A.~P.).

%%%%% References: %%%%%

%\newpage

%%%%% Figure captions %%%%%


\begin{thebibliography}{99}
\bibitem{Remk} Note that $\gamma $ is denoted as $P_{J}$ in Refs. \cite%
{OBO,BO,OB} and is called current spin polarization of FM-NS contact.
\end{thebibliography}
\end{document}